\documentclass{elsarticle}
 
\usepackage{graphicx, color}
\usepackage{amssymb}
\usepackage{amsmath}

\begin{document}

\begin{frontmatter}

\title{Real-time optical micro-manipulation using\\ optimized 
holograms generated on the GPU}
\author{S. Bianchi$^1$ and R. Di Leonardo $^2$}
\address{$^1$ Dipartimento di Fisica, Sapienza Universit\`a di Roma, I-00185, Roma, Italy}
\address{$^2$ CNR-IPCF, Dipartimento di Fisica, Sapienza Universit\`a di Roma, I-00185, Roma, Italy}

\begin{abstract}

Holographic optical tweezers allow the three dimensional, dynamic, multipoint
manipulation of micron sized objects using laser light.  Exploiting the massive
parallel architecture of modern GPUs we can generate highly optimized holograms
at video frame rate allowing the precise interactive micro-manipulation of
3D structures. 

\end{abstract}


\end{frontmatter}

\section{Introduction} 

Holographic optical tweezers (HOT) use light to manipulate matter at the micron
scale \cite{grier}. Dielectric objects, whose refractive index is higher than
the surrounding medium can be trapped in regions of high light intensity by
electromagnetic forces arising from the scattering of light \cite{ashkin}. To
achieve stable trapping in three dimensions, light has to be strongly focused
using a microscope objective with high numerical aperture. Many objects can be
trapped simultaneously if more than a single focal spot is generated around the
objective's focal plane.  Digital holography provides a way to achieve this by
applying a computer generated phase mask to a laser beam before it is sent
through the microscope objective. The commercial availability of Spatial Light
Modulators (SLM) has made this task easier by providing a reconfigurable
support for computer generated holograms which is connected to a PC through the
video output (usually DVI) on a standard video card  \cite{reicherter,
liesener, dufresne, curtis}.  The task of finding a phase mask that efficiently
redistributes the available laser power among an array of target focal spots is
not a straightforward one. Phase only  modulation can easily give rise to
unwanted focal spots (``ghost traps'') or large intensity variations. We
recently proposed an iterative procedure that achieves optimal efficiency and
uniformity in a few tens of steps \cite{rdl}. However the resulting
computational load is so high that the use of optimized algorithms for dynamic
manipulation is limited to those circumstances when the sequence of moves is
known in advance and holograms can be then pre-calculated. Such a slowness is
often considered as one of the major factors for preferring scanning beam
techniques \cite{aod} over digital holography for real-time applications. 

In this paper we demonstrate that CUDA \cite{cudaapp} enabled GPUs can generate
highly optimized holograms at a frame-rate that is fast enough to allow
interactive micro-manipulation using strong and uniform trap arrays.

\section{GPU device architecture}\label{GPU}

Graphic Processing Units (GPU) have brought the power of parallel calculus to
personal computers.  The possibility of using a personal computer to easily and
cheaply achieve the performances of an expensive CPU cluster is revolutionizing
computational physics in a wide range of fields including molecular dynamics
\cite{Liu}, Monte Carlo simulations \cite{preis}, finite element analysis
\cite{elsen}, lattice QCD \cite{qcd}.  The Compute Unified Device Architecture
(CUDA) is a general purpose parallel computing architecture introduced  by
NVIDIA.  CUDA provides a parallel programming model and software environment
allowing to exploit the massive parallel architecture of modern Graphic
Processing Units (GPU) for non-graphics applications.  General purpose parallel
algorithms can be implemented on a CUDA enabled GPU using a small set of C
extensions provided by the CUDA SDK.  The CUDA programming model closely
reflects the GPU hardware architecture.  A CUDA enabled GPU is composed of a
global memory and a variable number of multiprocessors. Each multiprocessor
includes eight scalar processor cores, two special function units, 8192
registers, a multithreaded instruction unit and one on-chip shared memory.  As
a result hundreds of cores can collectively run thousands of computing threads
that can share data without sending it over the system memory bus.  Threads are
arranged in a grid of blocks and each block is assigned to a multiprocessor.
In this way threads that belong to the same block can be synchronized and can
cooperate using shared memory.  Within a block, threads are arranged in groups
of 32 called warps, threads in a warp are physically executed in parallel and
are synchronized. Multiprocessors can only execute one warp at time, however if
threads in a warp are waiting to access global memory the multiprocessor can
stop executing that warp and switch to another one eliminating memory latency
time. 

Such an execution model requires specific optimization strategies that, for the
purpose of the present work, can be summarized in three general rules:

\begin{itemize}
\item[Rule A)] {\it Keep multiprocessors busy and hide memory latency.}

To this aim one should:
\begin{enumerate}
\item Group threads in a number of blocks that is multiple of the number of multiprocessors.
\item Choose the number of threads per block as a multiple of 32 to avoid wasting time with unfilled warps.
\item Maximize the number of active warps by using many threads per block.
\item When possible, avoid using conditional instructions that serialize the execution of a warp.
\end{enumerate}

\item[Rule B)] {\it Minimize read/write operations on global memory.}

Writing and reading global memory is very slow and sometimes it can be even
better to recalculate than to cache data. Shared memory must be used whenever
it can reduce the access to global memory. Shared memory is hundreds of times
faster than global memory but only 16k are currently available to any
multiprocessor.

\item[Rule C)] {\it Access global memory with coalesced calls.}

When all threads in a half warp execute a read/write instruction, the hardware
detects whether threads access consecutive global memory locations and
coalesces all these accesses. 

\end{itemize}

\section{Optimized algorithms for holographic trapping}
\begin{figure}[ht]
\centering
\includegraphics[width=.75\textwidth]{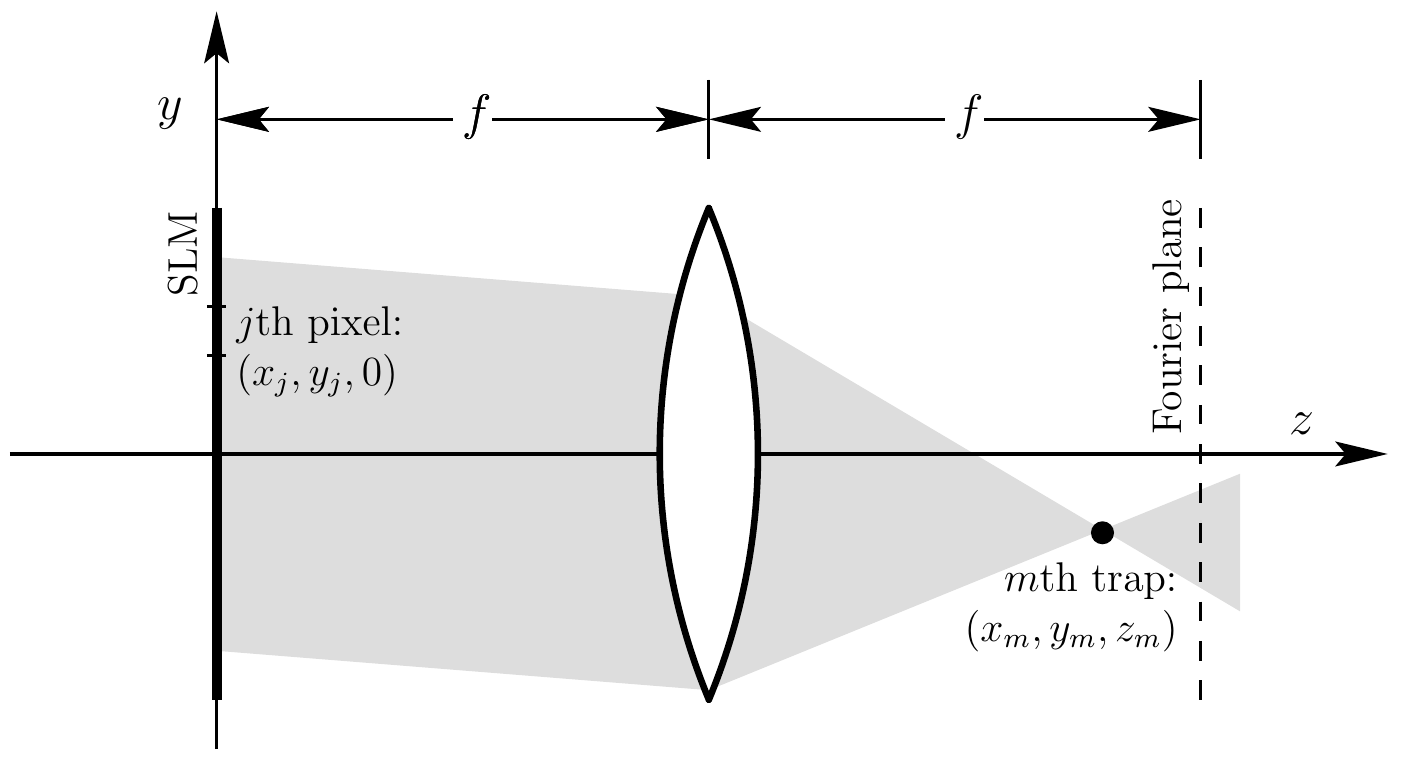}
\caption{Schematic representation of Fourier optics propagation from SLM plane (back focal
plane) to the front focal plane of the optical system.}
\label{fig:fourier}
\end{figure}

In back focal plane phase modulation we use an SLM to apply an array of phase
shifts to a plane wave at the back focal plane of a focusing optical system
(Fig. \ref{fig:fourier}).  Our task here is to calculate the best phase mask so
that the modulated wavefront propagating through the optical system is focused
onto an array of chosen target spots.
Given the phase shift on each pixel $\phi_j$ the complex field on a target
point $m$, with coordinates $x_m, y_m, z_m$, is given by \cite{rdl}:

\begin{equation}\label{V}
V_m=\frac{1}{N}\sum_{j=1,N}e^{i(\phi_j-\Delta_j^m)}
\end{equation}
\begin{figure}[h,t] \centering \includegraphics[width=.75\textwidth]{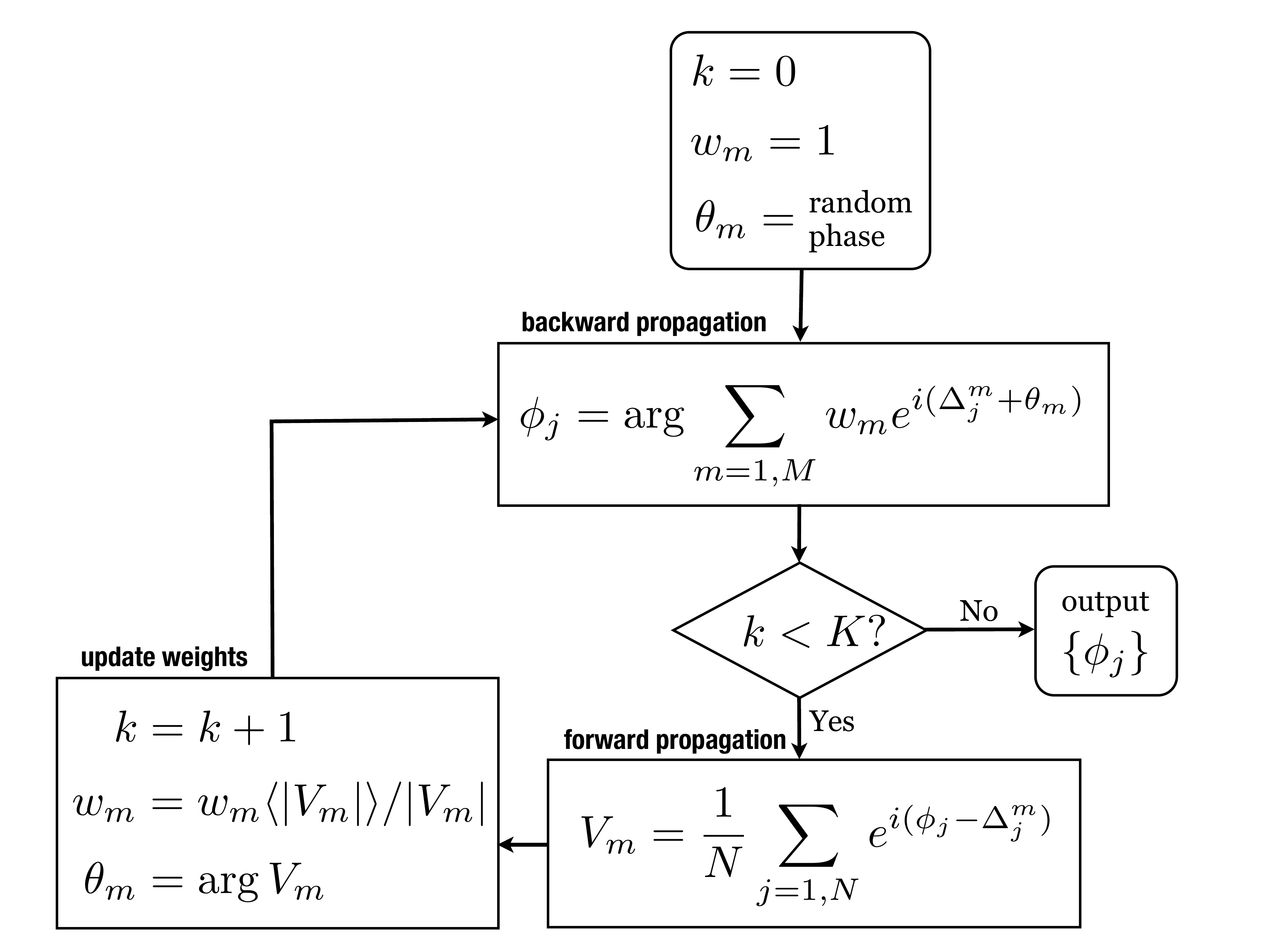}
\caption{Flowchart representing $K$ iterations of GSW algorithm.} \label{GSW}
\end{figure}

\begin{figure}[ht]
\centering
\includegraphics[width=.75\textwidth]{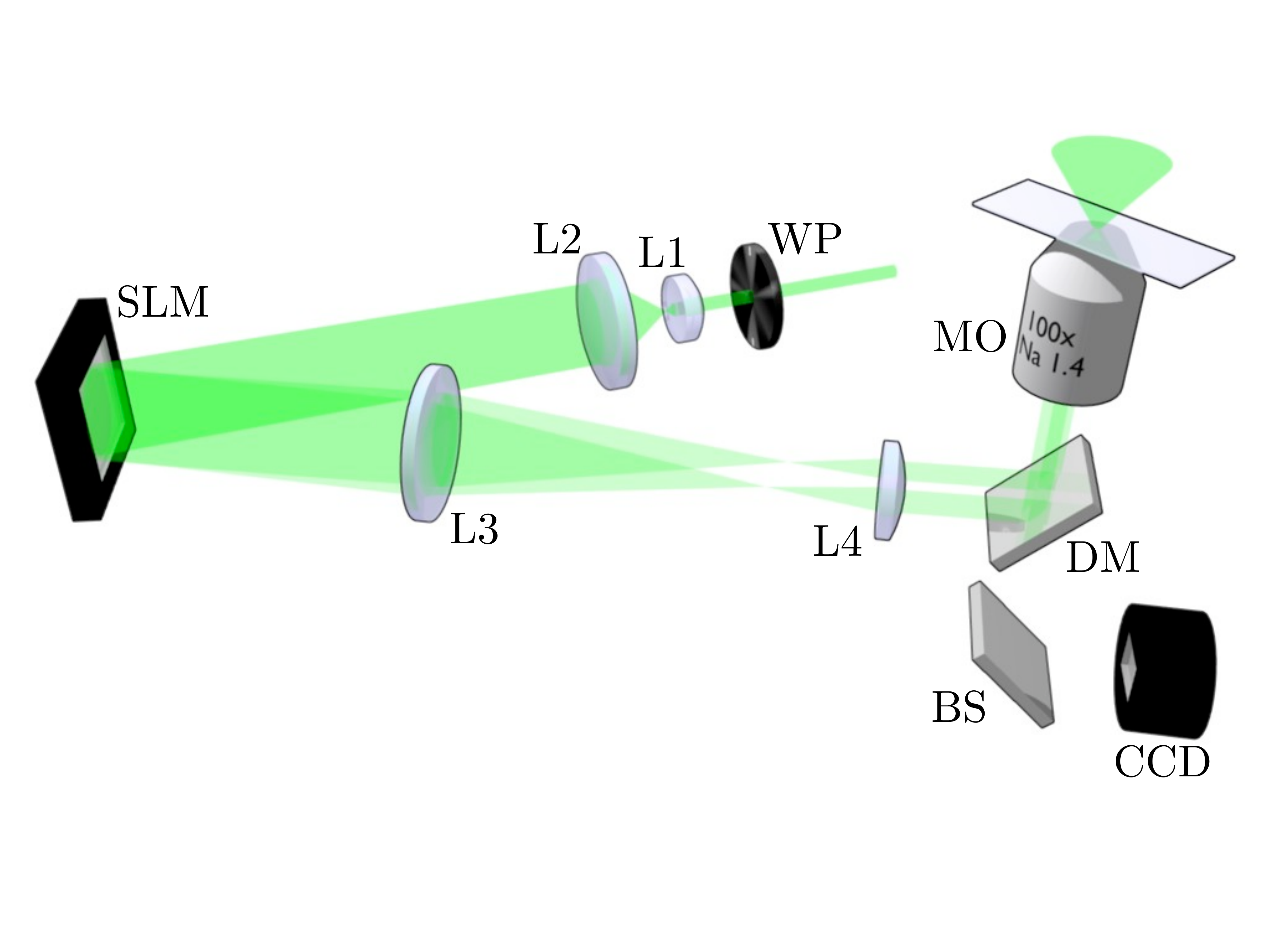}
\caption{
Schematic view of the experimental setup for holographic optical trapping.
L1,L2,L3,L4: planoconvex achromats, DM: dichroic mirror, BS: beam splitter, MO:
microscope objective, WP: half waveplate.
}
\label{fig:setup}
\end{figure}

Where N is the number of pixels, $i$ is the imaginary unit and $\Delta_j^m$ is
the phase acquired upon propagation:

\begin{equation}\label{Delta}
\Delta_j^m=\frac{z_m\pi}{\lambda f^2}(x_j^2+y_j^2)+\frac{2\pi}{\lambda f}(x_jx_m+y_jy_m)
\end{equation}

where $f$ is the effective focal length of the focusing optics (L3, L4, MO in
Fig. \ref{fig:setup}), $\lambda$ is the laser wavelength and $x_j, y_j$ are the
coordinates of the $j^{th}$ pixel. If we want to send all the light through a
single point $m=1$ then we should set $\phi_j=\Delta_j^1$, so that $V_1=1$.
When considering multiple traps, a phase only modulation might not be able to
split all the available power uniformly among the target points.  For each
pixel we now have the multiple choices $\Delta_j^m$ (the single trap holograms)
and finding a compromise could seem a hopeless task.  A first, reasonably fast
recipe is that of taking the complex superposition of single trap holograms
\cite{lesem}:
\begin{equation}\label{SR} \phi_j=\arg{\sum_{m=1,M} e^{i(\Delta_j^m+\theta_m)}}
\end{equation}
Where $\phi_j$ is again the phase of the $j^{th}$ SLM's pixel, $m^{th}$ is the
trap index, M is the number of traps, $\theta_m$ is a random phase relative to
the $m^{th}$ trap. Such a procedure, usually referred as the random
superposition algorithm (SR), is computationally rather fast but usually
results in ghost traps and poor uniformities, especially when dealing with
ordered structures.  A quantitative measure of the hologram performance can be
obtained by defining an efficiency ($e$) and a uniformity ($u$) parameters as a
function of the fractions of total power flowing through the $m^{th}$ trap
$I_m=|V_m|^2$:
\begin{equation}
e=\sum_m I_m\;,\;\;\;\;\;\;
u=1-\frac{\max[I_m]-\min[I_m]}{\max[I_m]+\min[I_m]}\;,\;\;\;\;\;\;
\end{equation}
A poor performance may result in particles getting trapped in unwanted ghost
trap sites or bead escape from temporary low intensity traps. When such events
are acceptable SR provides a good choice for real time manipulation, but if a
higher degree of control is required a more performing algorithm is needed. A
good candidate is the GSW algorithm (weighted Gerchberg-Saxton \cite{rdl})
which gives excellent results in terms of efficiency and uniformity. The basic
idea behind GSW is that, if aiming at uniform trap intensities with SR leads to
nonuniformities, we may hope that there's a choice of non uniform target traps'
intensities resulting in an evenly spread trapping light. GSW allows to
calculate such non uniform weights $w_m$ by the iterative procedure illustrated
in the flowchart reported in Fig. \ref{GSW}.
Angle brackets in the {\it update weights} box of Fig. \ref{GSW} represent averaging
over the trap index $m$.  After a few tens of iterations the procedure
converges to almost perfectly uniform trap intensities so that $|V_m|\simeq
\langle|V_m|\rangle$ and the weights $w_m$ don't get updated anymore.

\section{Implementing HOT algorithms on a CUDA device} 
\begin{figure}[h,t] \centering \includegraphics[width=.8\textwidth]{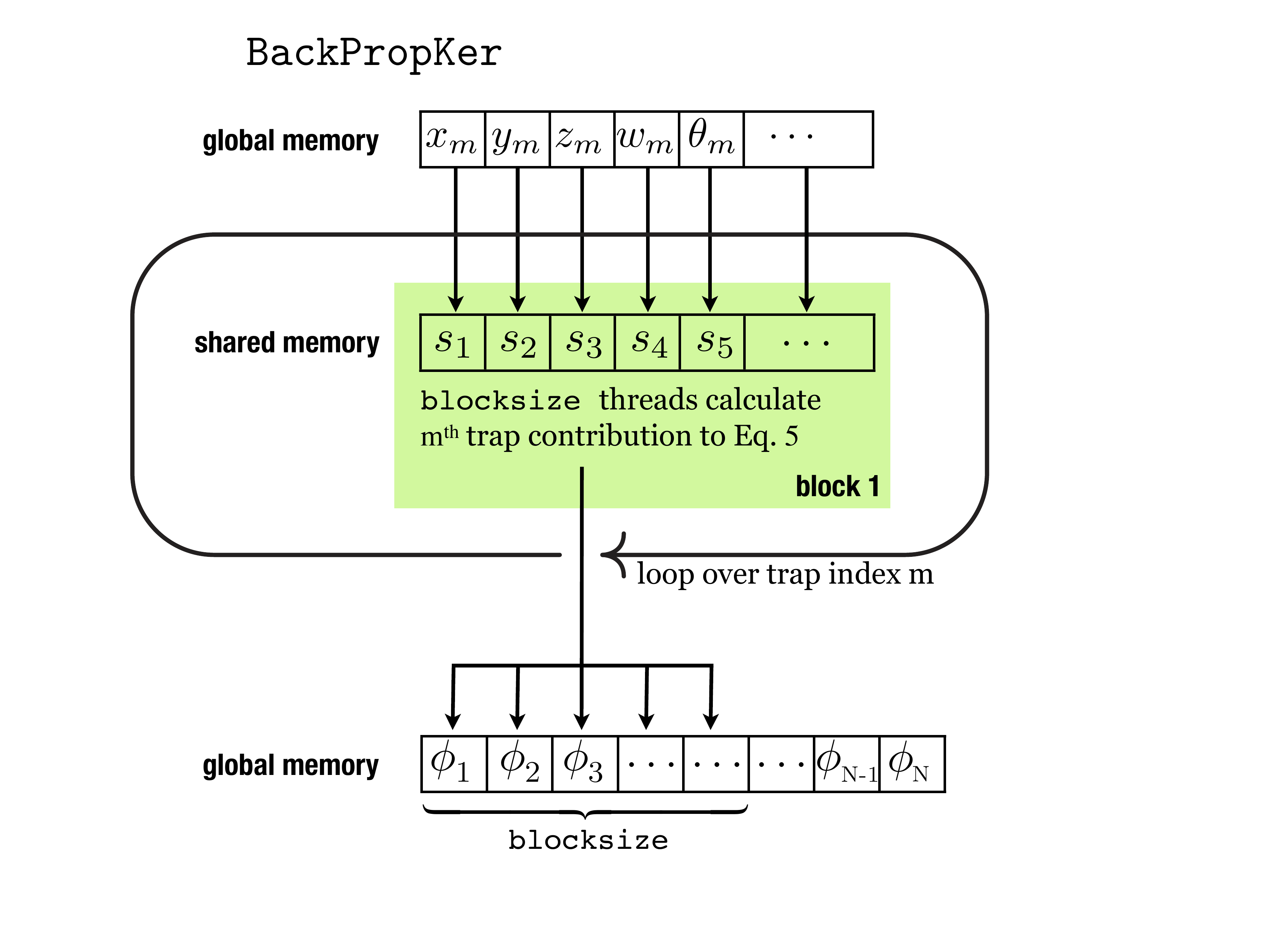}
\caption{Schematic diagram of the kernel {\ttfamily BackPropKer} calculating
the phase shifts on the SLM by a backward propagation of light emerging from the target spots.
} \label{BackPropKer} \end{figure}

The parallel architecture of GPUs is particularly suited for digital
holography, whose basic task is that of performing complex algebra over a large
array of independent pixels. In the field of digital holography GPUs have been
used for real-time holographic microscopy \cite{shimo, cheong} or holographic
displays \cite{masuda,Ahrenberg}. In the field of optical trapping, the
possibility of generating holograms with real-time frame rate is very
attractive for interactive applications. Early attempts always suffered the
slowness of CPU resulting in either slow or low efficiency holograms \cite{jon,
estela}.  More recently, custom shading programs running on the GPU have been
used to achieve a considerable speedup in hologram generation, although always
being limited to quick and poorly performing algorithms \cite{reich, preece}.
The CUDA architecture makes it a lot easier to implement more complex
algorithms in a general purpose environment which is not limited to graphic
applications.  When using a CUDA enabled video card, results can be also
computed directly on the frame buffer avoiding useless memory transfers.

\begin{table}[h]
\begin{tabular}{c c c c c}          
\hline
Rule A$^1$ & Rule B & Rule C$^2$ & t/trap (ms) & Speedup \\ [0.5ex]
\hline                       
Yes & No  & No  & 1.22 & 100\\              
Yes & Yes & No  & 0.42 & 290\\
No  & Yes & Yes & 0.47 & 260\\
Yes & Yes & Yes & 0.35 & 350\\ [1ex]    
\hline\\[-2mm]                            
\multicolumn{5}{l}{\footnotesize $^1$
Yes: {\ttfamily blocksize}=$32\times12=384$,
No: {\ttfamily blocksize}=$16$
}\\
\multicolumn{5}{l}{\footnotesize $^2$
Yes: threads in a block access SLM pixels as a linear array}\\[-1mm]
\multicolumn{5}{l}{\footnotesize
\hspace{2mm} No: threads in a block operate on square submatrices of SLM}
\end{tabular}
\label{table:nonlin}        
\caption{
Comparison table summarizing time costs of SR algorithm and the relative
importance of optimization rules. The GPU is a GeForce GTX 260 and
speedup data are evaluated with respect to a Pentium D 3.2 GHz.
}
\end{table}

Both of the previously discussed algorithms require the common step of backward
light propagation from the M target traps back to the N SLM pixels. In our case
the SLM is placed in the back Fourier plane of the optical system so backward
propagation is obtained by Eq. \ref{SR}. As shown in Fig \ref{BackPropKer},
the procedure can be translated into a kernel having as input arguments the
full trap structure described by the M coordinates, weights and phases: $(x_m,
y_m, z_m)$, $w_m$, $\theta_m$.  SR holograms are obtained by putting $w_m=1$
and choosing $\theta_m$ as random phases.  We implemented such a procedure in
the single kernel {\ttfamily BackPropKer} having a number of threads equal to
the number of SLM pixels. Each thread evaluates a single phase modulation
$\phi_j$ and stores it in a linear array residing in the global memory.
According to rule C in section \ref{GPU}, it is important that contiguous threads write on
contiguous pixels phase data so that coalesced memory access is guaranteed. As
discussed in rule A in section \ref{GPU} we use blocks containing a number of
threads that is large and multiple of 32. Each thread needs to access the full
trap structure so that a significant speedup can be achieved by preloading the
trap data in the shared memory as prescribed by rule B. In each block only M
threads cooperate to read the traps' data.
At this point we are ready to evaluate the time performance of {\ttfamily
BackPropKer} in generating holograms using the SR algorithm.  To this aim we
first generate M random phases $\theta_m$ on the CPU and than store the trap
structure on the global memory.  Using a GeForce GTX 260 we can generate
768$\times$768  SR holograms 350 times faster than using a Pentium D 3.2 GHz.
The time spent by SR to compute a hologram grows linearly with the number of
traps with a time per trap coefficient of 0.35 ms/trap. As an illustration of
the relative importance of the considered optimization rules, we compare in
Table 1 the most efficient kernel, where all this rules are obeyed, to
partially optimized kernels.

\begin{figure}[t]
\centering
\includegraphics[width=.8\textwidth]{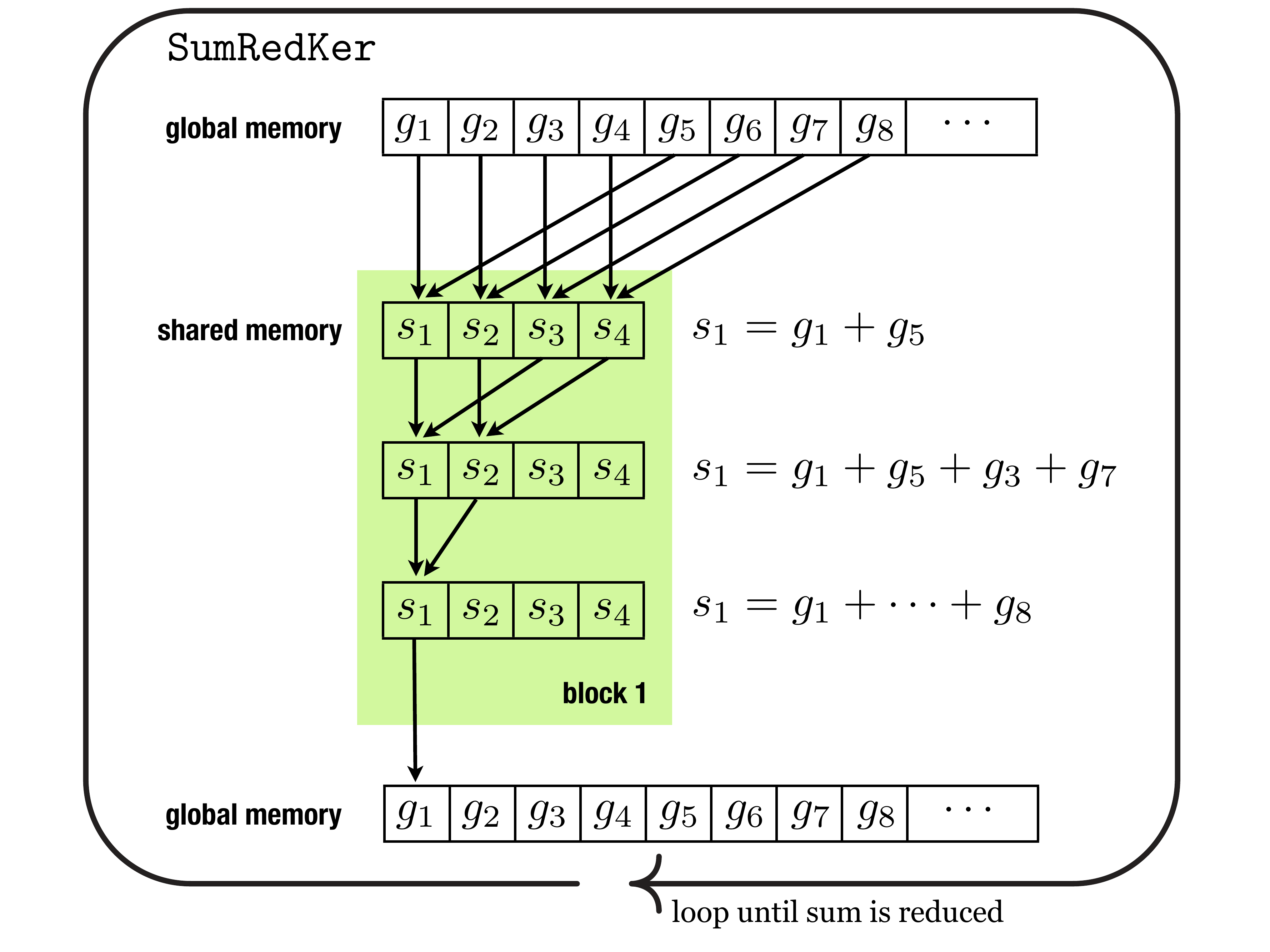}
\caption{
Schematic diagram of the kernel {\ttfamily SumRedKer}. 
For the sake of simplicity we only show a single block of
four threads. In the first step each thread sums up two elements of the
global memory and stores the result on shared memory. In the following steps
the number of active threads and the terms to sum are halved until a block is left
with one single term. The $i^{th}$ block will
write its partial sum on the $i^{th}$ address the global memory array.
}
\label{SRK}
\end{figure}

Turning now to the better performing GSW algorithm, in addition to a back
propagation kernel we need a procedure to forward propagate the fields from SLM
pixels to target traps (Eq. \ref{V}). Such a procedure can be decomposed into
two main tasks: i) calculate the contribution of each pixel to the complex
field $V_m$ on the $m^{th}$ trap's location, ii) sum up all contribution to
obtain $V_m$. The second task is a very common one and it's widely discussed in
the CUDA SDK examples. This procedure is based on the sum reduction kernel
{\ttfamily SumRedKer} that performs partial sums, reducing the number of terms.
A loop iterates {\ttfamily SumRedKer} until one single term is left containing
the sum of all elements.  A schematic representation of {\ttfamily SumRedKer}
is reported in Fig. \ref{SRK} where a single block is shown.  Each block
contains {\ttfamily blocksize} threads that perform the partial sum of
{\ttfamily 2$\times$blocksize} elements and writes the result back to the global
memory.  At the end of {\ttfamily SumRedKer} a number of terms equal to the
number of used blocks still remains to be summed.  Therefore a sequence of
$\log(N)/\log(${\ttfamily blocksize}$)$ kernels is needed to perform the whole
sum.

\begin{figure}[h]
\centering
\includegraphics[width=.75\textwidth]{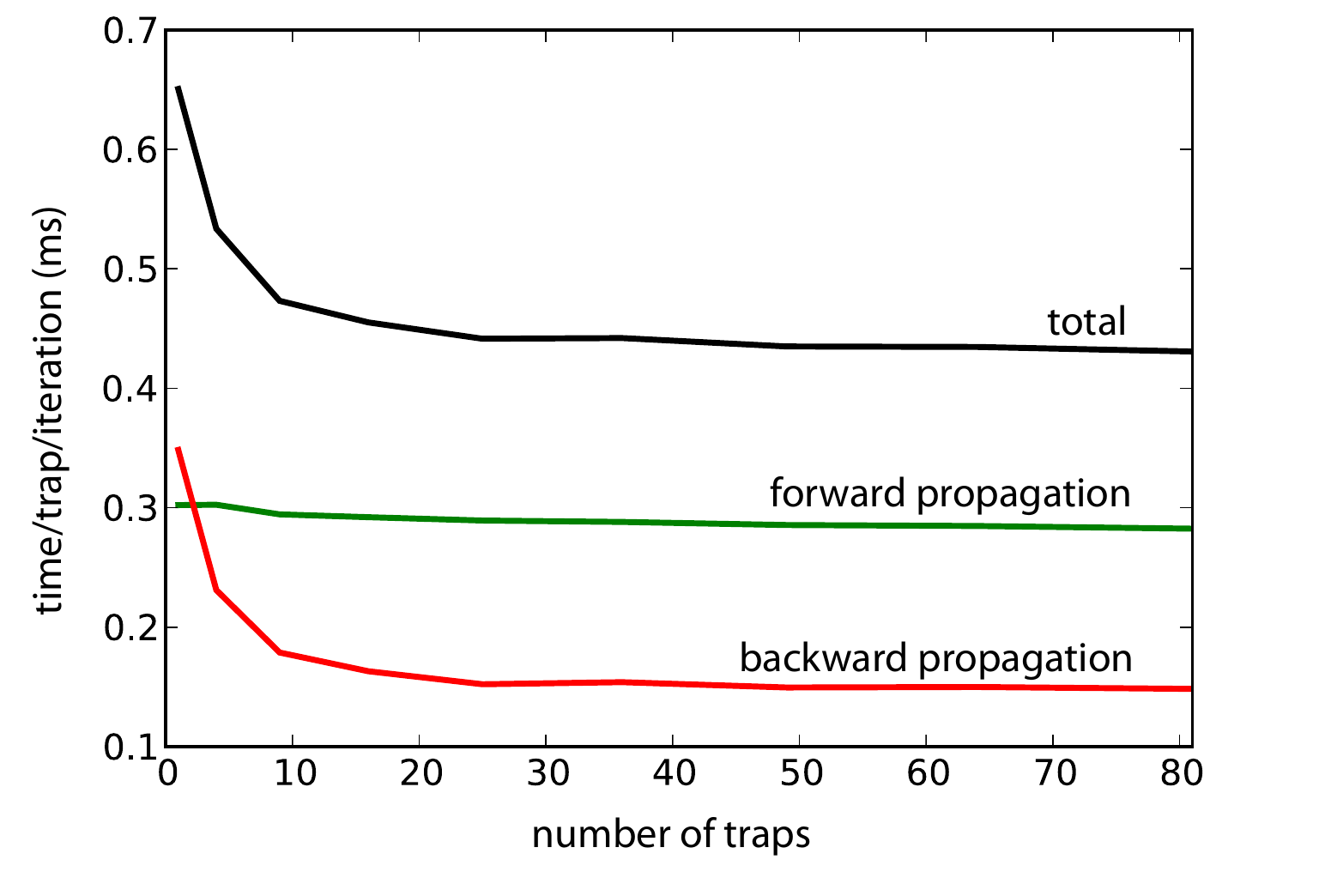}
\caption{
GSW computational time. The computational time per trap per iteration is reported as a function of the number of target points.
Total time (black line) is the sum of the time required for propagation (green line) plus the pixel phase evaluation (red line).
}
\label{fig:test}
\end{figure}
The evaluation of Eq. \ref{V} also requires the task of calculating the
contribution of the field radiating from each pixel to the total trap field
$V_m$.  Such a contribution is obtained calculating the phase shifts
$\Delta_j^m$ in Eq.  \ref{Delta} and building the complex exponentials $e^{i
(\phi_j-\Delta_j^m)}$.  In order to reduce read/write operations on global
memory we use a slightly different version of the sum reduction kernel as the
first partial sum step.  The first {\ttfamily SumRedKer} will begin having the
phase modulations $\phi_j$ on the global memory locations $g_j$ in Fig.
\ref{SRK} so that we need to calculate complex exponential before the first
write on shared memory (i.e.  $s_1= e^{i(\phi_1-\Delta_1^m)}+
e^{i(\phi_5-\Delta_5^m)}$).  The phases $\Delta_j^m$ (M*N in total) are needed
both for forward and backward propagation routines. Observing that such phases
are fixed for a chosen trap geometry, one could think that precaching them in
global memory could save computational time. However we checked that direct
calculation is always faster (see rule B).  Once $V_m$s are known, the
calculation of the weights $w_m$ is quick and straightforward.  The time
required by GSW grows almost linearly with the number of traps or iterations.
In Fig. \ref{fig:test} we report the computational time per trap per iteration as a function
of traps number. Deviations from linearity are observed for small traps number
evidencing the presence of a time cost which is essentially independent from
the number of traps and is probably due to memory read/write operations.  As we
can see from the figure, we can neglect the small deviations from linearity and
define a time per trap per iteration.
Using a GeForce GTX 260 we obtain 0.44 ms/trap/iteration obtaining a 45x speedup
respect to a Pentium D 3.2 GHz.

\section{Real-time manipulation}

\begin{figure}[ht]
\centering
\includegraphics[width=.75\textwidth]{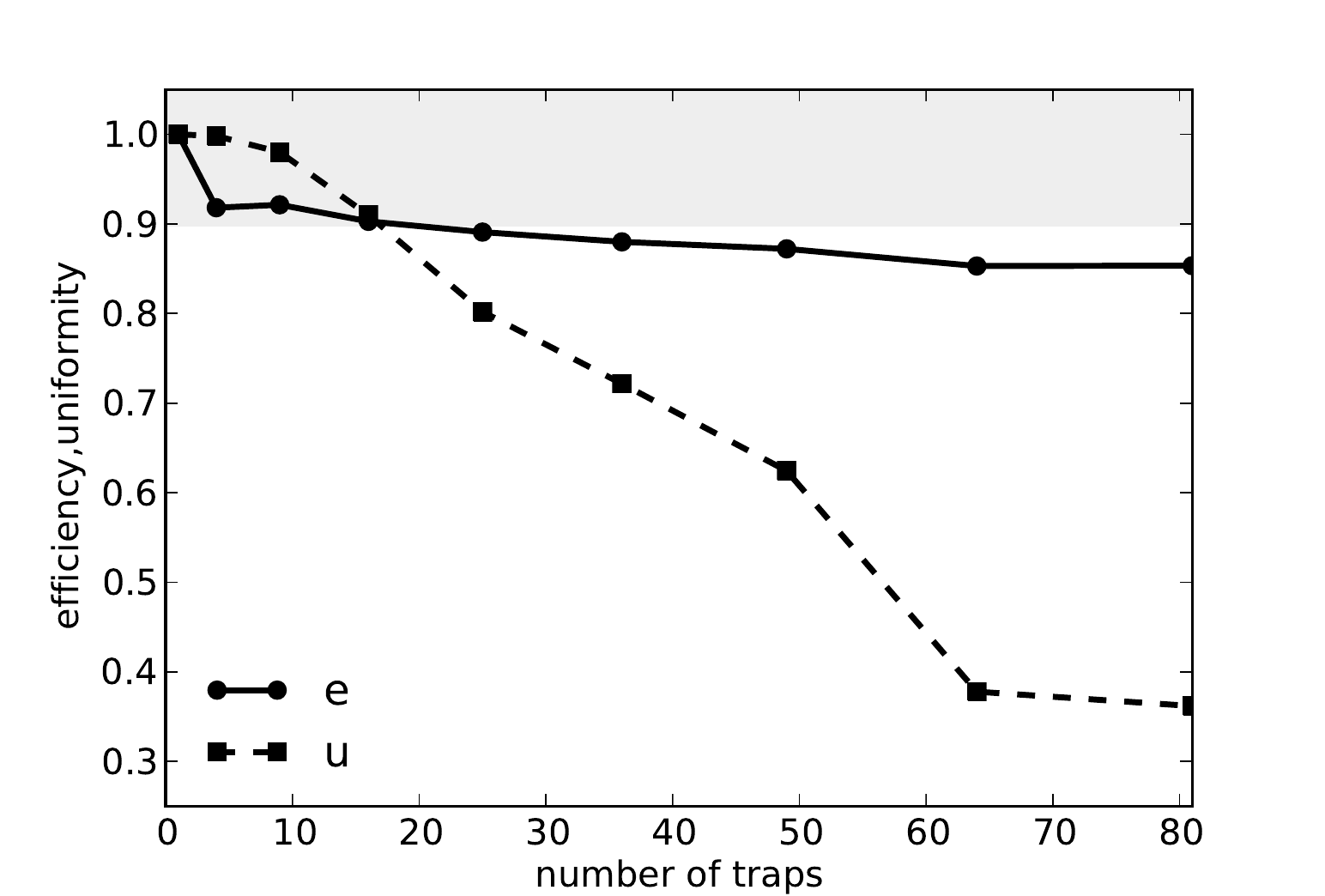}
\caption{
GSW performance. We report the efficiency ($e$) and uniformity ($u$) for GSW generated holograms as a function of the number of traps. The number of GSW iterations is always such to work at a fixed framerate of 20 Hz. Holograms with a performance above 90\% can be generated at 20Hz for trap arrays as large as 16.
}
\label{fig:eu}
\end{figure}

Our optical tweezers are based upon a Nikon TE2000U inverted microscope with a
100x objective lens, NA 1.4. To form the trap we use a Nd:YAG laser,
frequency-doubled to give a maximum power of 3 W at 532 nm (LaserQuantum Opus).
After expansion and collimation, the beam from this laser is reflected off a
computer-controlled SLM (HoloEye LCR 2500).  Our SLM is based on a liquid
crystal reflective micro-display. A laser beam reflecting off the SLM will
emerge with a phase retardation that can be modulated on a pixel by pixel
basis. Phase modulation is achieved by electrically addressing the pixels and
therefore locally reorienting the nematic axis of liquid crystal molecules.
When a grayscale, 8bit depth image is displayed on the SLM, a proper pattern of
voltages is relayed to the pixels so that each grayscale value is linearly
mapped to a phase shift ranging from 0 to 2$\pi$. Light reflected off the SLM
is then coupled to the microscope by projecting a demagnified image of the SLM
plane on the back focal plane of the microscope objective. An array of optical
traps is then produced around the front focal plane of the objective located in
a colloidal water suspension above the coverslip. The SLM was controlled by a
host PC equipped with a NVidia GeForce GTX 260 video card. User input is
managed by a GUI mainloop thread (Tkinter) running in a Python shell while a
Python module wraps the CUDA library functions providing a high level interface
to the GPU hologram generation. 

\begin{figure}[ht]
\centering
\includegraphics[width=.75\textwidth]{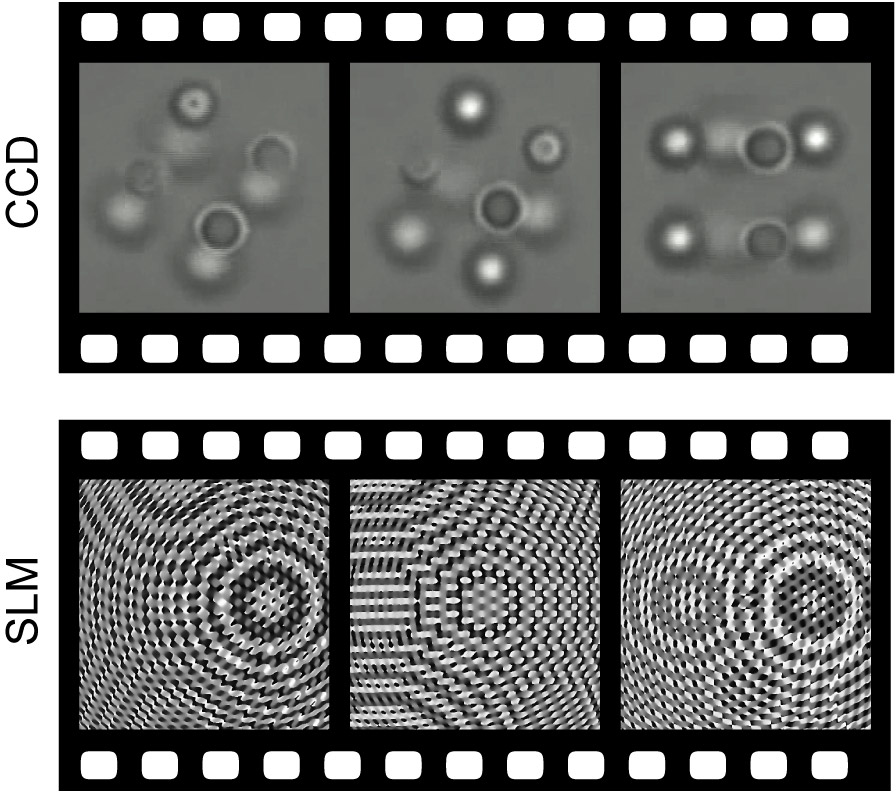}
\caption{
Frames from a movie showing the interactive micro-manipulation of 8 silica beads with 2 $\mu$m diameter (watch the full movie in supplementary online material). The beads are arranged on the vertices of a 5 $\mu$m side cube which is then rigidly rotated.
Bottom row shows the corresponding frames (holograms) displayed on the SLM.
}
\label{fig:frames}
\end{figure}

As a demonstration of real-time manipulation using optimized GPU generated
holograms, we show the simultaneous trapping and manipulation of  eight silica
beads (2$\mu$m diameter) in water. Optimized holograms are obtained with 5 GSW
iterations at a rate of  48 Hz following user input.  Fig. \ref{fig:frames}
shows three frames from the corresponding SLM and CCD timelines. While a
hologram movie  is displayed on the SLM (lower timeline) based on user input, a
dynamic 3D micro-hologram, consisting of an array of moving bright light spots,
is projected in the sample volume providing dynamical, and real-time
reconfigurable optical traps. Trapped beads are imaged with bright light
illumination on a CCD camera (upper timeline). The actual frame-rate is
slightly lowered due to time lost in copying from device memory to host memory
and then back to the video card output where the SLM is attached. This further
delay could be avoided exploiting CUDA-OpenGL interoperability. In this way
holograms could be calculated directly on the frame buffer and displayed on the
SLM without passing through the host. Ultimately the frame-rate is limited by
SLM response time, which, for liquid crystal based devices, is typically about
20 Hz. Such a frame-rate allows to perform a large enough number of GSW
iterations to generate large arrays of traps with a high efficiency and
uniformity.  In Fig.\ref{fig:eu} we report the efficiency ($e$) and uniformity
($u$) for GSW generated holograms as a function of the number of traps arranged
in a 2D square grid.  The number of GSW iterations is always such that
holograms are generated at a fixed framerate of 20 Hz.  Though efficiency is
never lower than 85\% uniformity falls down to 0.36 for a 9x9 grid where only
one GSW iterations is allowed in order to keep the frame-rate at 20 Hz.  We
note here that even one single GSW iteration results in a significant
improvement in performance over SR which would give an efficiency of 70\% and a
uniformity of only a few percents.

In conclusion, we have used a CUDA enabled video card to generate optimized
holograms for optical trapping with a speedup of 350x (SR) and 45x (GSW) over
the host CPU. The obtained speedup allowed us to trap and manipulate
multiparticle 3D structures with efficient and uniform trap arrays in real
time. Our results demonstrate that the high computational load of hologram
generation cannot be considered any longer as a limiting factor of holographic
trapping for real time applications.  We acknowledge support from INFM through
the Seed-project.


\begin{thebibliography}{00}

\bibitem{grier} 
D.G. Grier, 
A revolution in optical manipulation, 
Nature 424 (2003) 810-816.

\bibitem{ashkin} A. Ashkin, J. M. Dziedzic, J. E. Bjorkholm, 
Observation of a single-beam gradient force optical trap for dielectric particles,
Opt. Lett. 11 (1986) 288-290.

\bibitem{reicherter} M. Reicherter, T. Haist, E.U. Wagemann, H.J. Tiziani, 
Optical particle trapping with computer-generated holograms written on a liquid-crystal display,
Opt. Lett. 24 (1999) 608-610.

\bibitem{liesener} J. Liesener, M. Reicherter, T. Haist, H.J. Tiziani,
Multi-functional optical tweezers using computer-generated holograms,
Opt. Commun. 185 (2000) 77-82. 

\bibitem{dufresne} E.R. Dufresne, G.C. Spalding, M.T. Dearing, S.A. Sheets, D.G. Grier,
Computer-generated holographic optical tweezers arrays,
Rev. Sci. Instrum. 72 (2001) 1810-1816.

\bibitem{curtis} J. Curtis, B.A. Koss, D.G. Grier,
Dynamic holographic optical tweezers, 
Opt. Commun. 207 (2002) 169-175.

\bibitem{rdl}
R. Di Leonardo, F. Ianni, G. Ruocco,
Computer generation of optimal holograms for optical trap arrays,
Opt. Express 15 (2007) 1913-1922.
 
\bibitem{aod}
K. Visscher, G. J. Brakenhoff, and J. J. Kroll,
Micromanipulation by multiple optical traps  created by a single fast scanning
trap integrated with the bilateral confocal scanning laser microscope,
Cytometry 14 (1993) 105-114. 

\bibitem{cudaapp}
http://www.nvidia.com/object/cuda\_home.html

\bibitem{Liu}
W. Liu, B. Schmidt, G. Voss, and W. Müller-Wittig, 
Accelerating molecular dynamics simulations using Graphics Processing Units with CUDA, 
Comp. Phys. Comm. 179 (2008) 634-641.

\bibitem{preis}
T. Preis, P. Virnau, W. Paul, J.J. Schneider,
GPU accelerated Monte Carlo simulation of the 2D and 3D Ising model,
J. Comp. Phys. 228 (2009) 4468-4477.

\bibitem{elsen}
E. Elsen, P. LeGresley, E. Darve,
Large calculation of the flow over a hypersonic vehicle using a GPU,
J. Comp. Phys. 227 (2008) 10148-10161.

\bibitem{qcd}
G. I. Egri, Z. Fodor, C. Hoelbling, S.D. Katz, D. Nógrádi, K.K. Szabó,
Lattice QCD as a video game, Comp. Phys. Comm. 177 (2007) 631-639.

\bibitem{lesem} 
L.B. Lesem, P.M. Hirsch, J.A. Jordan,
The kinoform: a new wavefront reconstruction device,
IBM J. Res. Dev. 13 (1969) 150-155.

\bibitem{shimo}
T. Shimobaba, Y. Sato, J. Miura, M. Takenouchi and T. Ito,
Real-time digital holographic microscopy using the graphic processing unit,
Opt. Express 16 (2008) 11776-11780.

\bibitem{cheong}
F.C. Cheong, B. Sun, R. Dreyfus, J. Amato-Grill, K. Xiao, L. Dixon, D.G. Grier,
Flow visualization and ﬂow cytometry with holographic video microscopy,
Opt. Express 17 (2009) 13071-13079.


\bibitem{masuda}
N. Masuda, T. Ito, T. Tanaka, A. Shiraki, and T. Sugie,
Computer generated holography using a graphics processing unit,
Opt. Express 14 (2006) 603-608.

\bibitem{Ahrenberg}
L. Ahrenberg, P. Benzie, M. Magnor, and J. Watson, 
Computer generated holography using parallel commodity graphics hardware, 
Opt. Express 14 (2006) 7636-7641.

\bibitem{jon}
J. Leach et al., Interactive approach to optical tweezers control,
 Appl. Optics  45 (2006) 897-903.

\bibitem{estela}
E. Pleguezuelos, A. Carnicer, J. Andilla, E. Martin-Badosa and M. Montes-Usategui,
Fast generation of holographic optical tweezers by random mask encoding of Fourier components
Comp. Phys. Comm. 176 (2007) 701-709.

\bibitem{reich}
M. Reicherter, S. Zwick, T. Haist, C. Kohler, H. Tiziani, and W. Osten,
Fast digital hologram generation and adaptive force measurement in
liquid-crystal-display-based holographic tweezers, Appl. Opt. 45 (2006) 888–896. 

\bibitem{preece}
D. Preece, R. Bowman, A. Linnenberger, G. Gibson, S. Serati and M. Padgett, Increasing trap stiffness with position clamping in holographic optical tweezers, Opt. Express 17 (2009) 22718-22725.

\end{thebibliography}
\end{document}